\documentclass[
  journal=pasa,
  manuscript=Research-Paper,
  year=2025,
  volume=37,
]{cup-journal}

\usepackage{amsmath}
\usepackage{url}
\usepackage[nopatch]{microtype}
\usepackage{booktabs}
\usepackage{subcaption}
\usepackage{multirow}
\usepackage{array}
\usepackage{booktabs} 
\usepackage{adjustbox} 
\usepackage{natbib}
\newcolumntype{C}[1]{>{\centering\arraybackslash}p{#1}}

\title{Overlap Zoo Beta: A Catalogue of {\(\sim\)800} Occulting Pairs in the DESI Legacy Survey using Citizen Science}

\author{T. Butrum}
\affiliation{Department of Physics and Astronomy, University of Louisville, Louisville, 40208, Kentucky, United States}
\email[T. Butrum]{tbutrum@trevor-astronomy.com}

\author{B. Holwerda}
\affiliation{Department of Physics and Astronomy, University of Louisville, Louisville, 40208, Kentucky, United States}

\author{W. C. Keel}
\affiliation{Department of Physics and Astronomy, University of Alabama, 35487, Alabama, Tuscaloosa, United States}

\author{C. Robertson}
\affiliation{Department of Physics and Astronomy, University of Louisville, Louisville, 40208, Kentucky, United States}

\author{I. Castellano}
\affiliation{Department of Physics, Geology, and Engineering Technology, Northern Kentucky University, Highland Heights, KY 41099}

\author{S. Pandey}
\affiliation{Department of Physics and Astronomy, University of Louisville, Louisville, 40208, Kentucky, United States}

\author{S M R. Adnan}
\affiliation{Department of Physics and Astronomy, University of Louisville, Louisville, 40208, Kentucky, United States}

\author{L. C. Bills}
\affiliation{Department of Physics and Astronomy, University of Louisville, Louisville, 40208, Kentucky, United States}

\author{D. Patel}
\affiliation{Department of Physics and Astronomy, University of Louisville, Louisville, 40208, Kentucky, United States}

\author{K. Cook}
\affiliation{Department of Physics and Astronomy, University of Louisville, Louisville, 40208, Kentucky, United States}

\author{T. Hardin}
\affiliation{Department of Physics and Astronomy, University of Louisville, Louisville, 40208, Kentucky, United States}

\author{A. Palao}
\affiliation{Department of Physics and Astronomy, University of Louisville, Louisville, 40208, Kentucky, United States}

\author{B. Connelly}
\affiliation{Department of Physics and Astronomy, University of Louisville, Louisville, 40208, Kentucky, United States}

\author{M. Morton}
\affiliation{Department of Physics and Astronomy, University of Louisville, Louisville, 40208, Kentucky, United States}

\begin{raggedright}
\end{raggedright}

\begin{document}

\begin{abstract}
Overlapping galaxies, in which a foreground galaxy partially overlaps a background galaxy, offer a unique opportunity to measure dust attenuation, a key nuisance parameter in galaxy studies, empirically and in great detail by modelling the light of both the foreground and background galaxy and inferring the missing light in the overlapping region.  However, the current catalogue of overlapping pairs is relatively limited in number compared to catalogues dedicated to individual galaxies. Expanding this catalogue is not only a necessity to facilitate further detailed dust studies beyond the few limited studies conducted thus far, but also to improve pair-to-pair variance and support automated identification through machine learning techniques. To achieve this, we utilise galaxies classified as "overlapping" from Galaxy Zoo DECaLS (GZD-1, -2, and -5), along with images from Data Release 10 (DR10) of the DESI Legacy Survey, in our individual citizen science project to classify these pairs directly using volunteers. This new catalogue will not only provide a wealth of targets for future dust studies but will also contribute to a deeper understanding of these pairs and dust as a whole.
\end{abstract}

\section{Introduction}


Only a small fraction of the mass of the interstellar medium (ISM) in a galaxy consists of interstellar dust \textcolor{black}{\citep{Casasola19, La-Mura25}}. Yet, that minute fraction plays a significant role in our astronomical observations. Dust attenuation, which encompasses both the absorption and scattering of starlight as well as the geometry of the sources and the ISM itself, hinders accurate measurements of the intrinsic properties of galaxies by influencing and reshaping the spectral energy distribution (SED) \textcolor{black}{\citep{Salim20}}. \textcolor{black}{As a result, standard and fundamental galaxy evolution parameters, such as star-formation rates (SFRs) and stellar mass (\(M_\star\)), are subject to systematic uncertainties.}

\textcolor{black}{The simplest way to account for this wavelength-dependent attenuation of starlight by dust is to introduce a single attenuation curve that captures the complex blending of dust properties and source geometry \citep{Buat18}. However, developing a single, universal attenuation law remains an open challenge in the modern era. The differences in attenuation relations used between sight-lines in empirical studies (e.g., \citealp{Calzetti94, Calzetti00, Battisti16, Battisti17, Johnson07, Wild11}), IRX-\(\beta\) studies (e.g., \citealp{Meurer99, Seibert05, Casey14a}), and model-based studies (e.g, \citealp{Burgarella05, Conroy10, Leja17, Salim18a}) all demonstrate that a universal law does not currently exist. Some explanations of this diversity may involve star-dust geometric effects, metallicity, or sampling biases introduced by emission lines (e.g., \citealp{Calzetti94, Calzetti00, Charlot00, Gordon03, Wild11, Reddy15, Buat18}). However, we ultimately need a deeper understanding of interstellar dust and its influence on our observational data to be able to explain the cause of these differences confidently.}

Occulting pairs of galaxies, in which a foreground galaxy partially overlaps a background galaxy, are becoming increasingly valuable for studying dust attenuation and other dust properties in galaxies. These occulting systems present a unique opportunity to derive attenuation measurements by modelling the light from both galaxies and inferring the missing light in the overlapping regions, leaving only the light attenuated by dust remaining. This approach, first introduced by \cite{White92}, demonstrated practical use in many areas of photometry \citep{Andredakis92, Berlind97, Holwerda09hst, Holwerda13, Keel23, Robertson24a}. However, despite the success of this technique, we are still limited to a relatively small number of occulting pairs to which we can apply it and derive attenuation and other measurements.


Researchers have made several efforts to identify and catalogue occulting pairs. Currently, the master catalogue of occulting pairs consists of \textcolor{black}{\(\sim2,400\)} pairs suitable for dust studies. These pairs include 15 identified in Hubble Space Telescope images (HST; \citealp{Keel01a, Keel01b, Elmegreen01a, Holwerda09hst, Holwerda13}), and 2,079 identified in the Sloan Digital Sky Survey (SDSS) spectroscopic catalogue, with 86 pairs identified by \cite{Holwerda07hst} and 1,993 identified by \cite{Keel13} through forum science using a Galaxy Zoo project \citep{Lintott08}. More recently, 280 were identified in the Galaxy and Mass Assembly (GAMA) catalogue by \cite{Holwerda15}. 

Expanding this master catalogue will undoubtedly provide numerous benefits. As the sample size of this catalogue increases: (1) more 'ideal pairs'—a foreground face-on spiral partially overlapping a smooth background galaxy—will be available for follow-up studies, (2) the variance of the data will increase, providing a better ability to capture the true variability of pairs, thus improving the statistical robustness of measurements in the catalogue, and (3) it will pave the way for improving the accuracy of machine learning predictions.


The catalogue produced by \cite{Keel13} represented a significant advance in this effort, significantly increasing the number of known occulting pairs nearly 20 times. However, it relied solely only on a single independent expert's classifications, where volunteers helped identify potential occulting candidates, rather than providing direct classifications. Given the increasing success of citizen science projects in large-scale astronomical classifications, especially in identifying low-redshift galaxies, directly incorporating volunteer classifications into the workflow represents the next promising strategy for improving and facilitating occulting pair classification. Therefore, in this study, we introduce a method that attempts to seamlessly integrate a citizen science approach into the classification process of a Galaxy Zoo-like project, allowing for a more organised and scalable approach to classifying occulting pairs. Our foremost goal is to expand the sample size of our current catalogue of occulting pairs and to evaluate the effectiveness of our (relatively small-scale) citizen science approach in making these classifications in preparation for future higher resolution surveys such as Euclid \citep{EuclidCollab25}, SPHEREx \citep{Crill20}, and LSST \citep{Ivezic19}.


We organise this paper as follows. Section \ref{projectdesc} describes the project as a whole, including our selection and image generation process. Section \ref{data_reduc} details our data reduction steps taken before assembling the catalogue. Section \ref{catalogue} provides an overview of the main catalogue and its results. Section \ref{accuracy} evaluates the effectiveness of both the project and the resulting catalogue and compares our catalogue to previous catalogues. Section 
\ref{limitations} discusses the limitations found after concluding the project and outlines directions for future work. Finally, Section \ref{conclusion} concludes the paper by summarizing our main findings.


\section{{Project Description}}
\label{projectdesc}

Our citizen science project includes a sample of galaxies identified as "overlapping" \textcolor{black}{by human volunteers} in Galaxy Zoo DECaLS (GZD) Data Releases 1, 2, and 5 (GZD-1, GZD-2, and GZD-5; \citealp{Walmsley22}) and images from Data Release 10 (DR10) of the DESI Legacy Survey \citep{Dey19}. \textcolor{black}{DR10 was chosen as the source of images presented to volunteers because it offers improved depth and uniformity relative to the earlier GZD data releases, providing better data for visual classification.} In the following, we describe our selection process and the image generation process for the DR10 images presented to the classifiers. We also briefly detail the workflow and web-based interface presented to each volunteer during their classification process.

\subsection{Sample Selection}
\label{sampleselect}

\begin{figure}[!htb] 
\centering
\begin{subfigure}{\textwidth}
\graphicspath{ {Images/}}
\centering
\includegraphics[width=\textwidth]{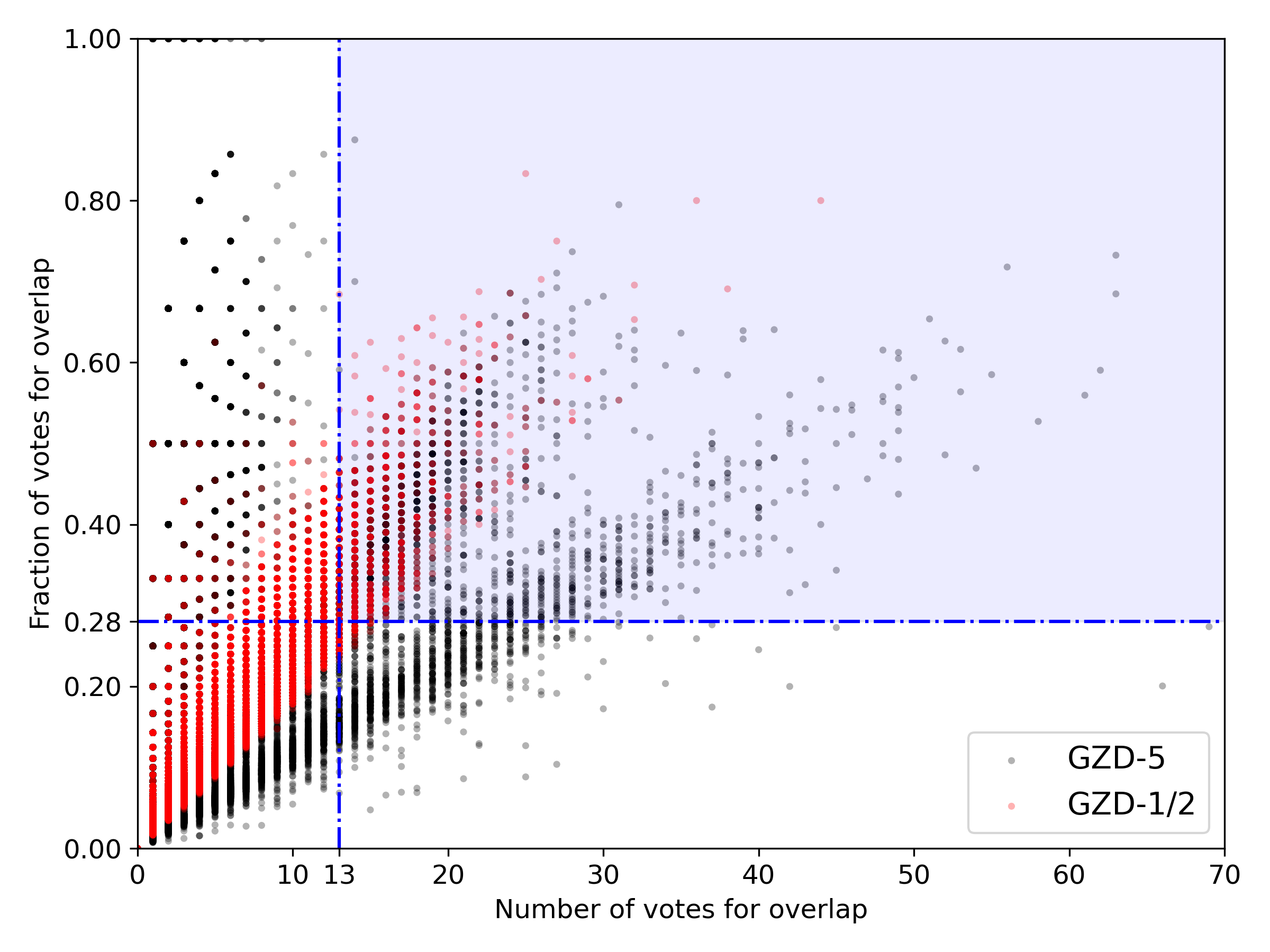}
\centering
\end{subfigure}%
\caption{The distribution of galaxies classified as 'overlapping' in GZD-1, -2, and -5. We plot the GZD-1 and GZD-2 data as black dots, the GZD-5 data as red dots, and our selections of the combined data as blue dots. Our selection criteria include only galaxies classified as overlapping with votes >13 and voting fractions >0.28.}
\label{selection}
\end{figure}

We selected candidate occulting pairs by applying targeted data cuts to galaxies voted by \textcolor{black}{human} volunteers as "overlapping" in the GZD datasets. The goal of our sample selection was to include only valid overlapping pairs while minimizing false overlaps, such as star-galaxy and interacting pairs. Since GZD provides only votes and voting fractions, we restricted our selection to these two variables. However, these restrictions presented a challenge, as many star-galaxy overlaps garnered high votes and voting fractions from volunteers in GZD-1, GZD-2, and GZD-5, with many classifications of actual overlaps consisting of low votes and voting fractions. Therefore, as a conservative measure, we required that all pairs in our selection have \textcolor{black}{"overlapping"} votes >13 and a voting fraction >0.28 (see Figure \ref{selection}). We note that it is well known that this selection process can be flawed, and our sample may contain many false classifications, such as star or merging or interacting overlaps \citep{Darg10a, Lintott11a}. Applying these restrictions and \textcolor{black}{manually removing 54 images due to noticeable visual artifacts} resulted in 2,380 occulting pairs \textcolor{black}{(\(1.56\%\) of the original GZD sample)} that met this criterion, which we regard as our "original" sample.

\subsection{Generating Images}
\label{images}

\begin{figure*}[!htbp] 
\centering
\begin{subfigure}{1\textwidth}
\graphicspath{ {Images/}}
\centering
\includegraphics[width=\textwidth]{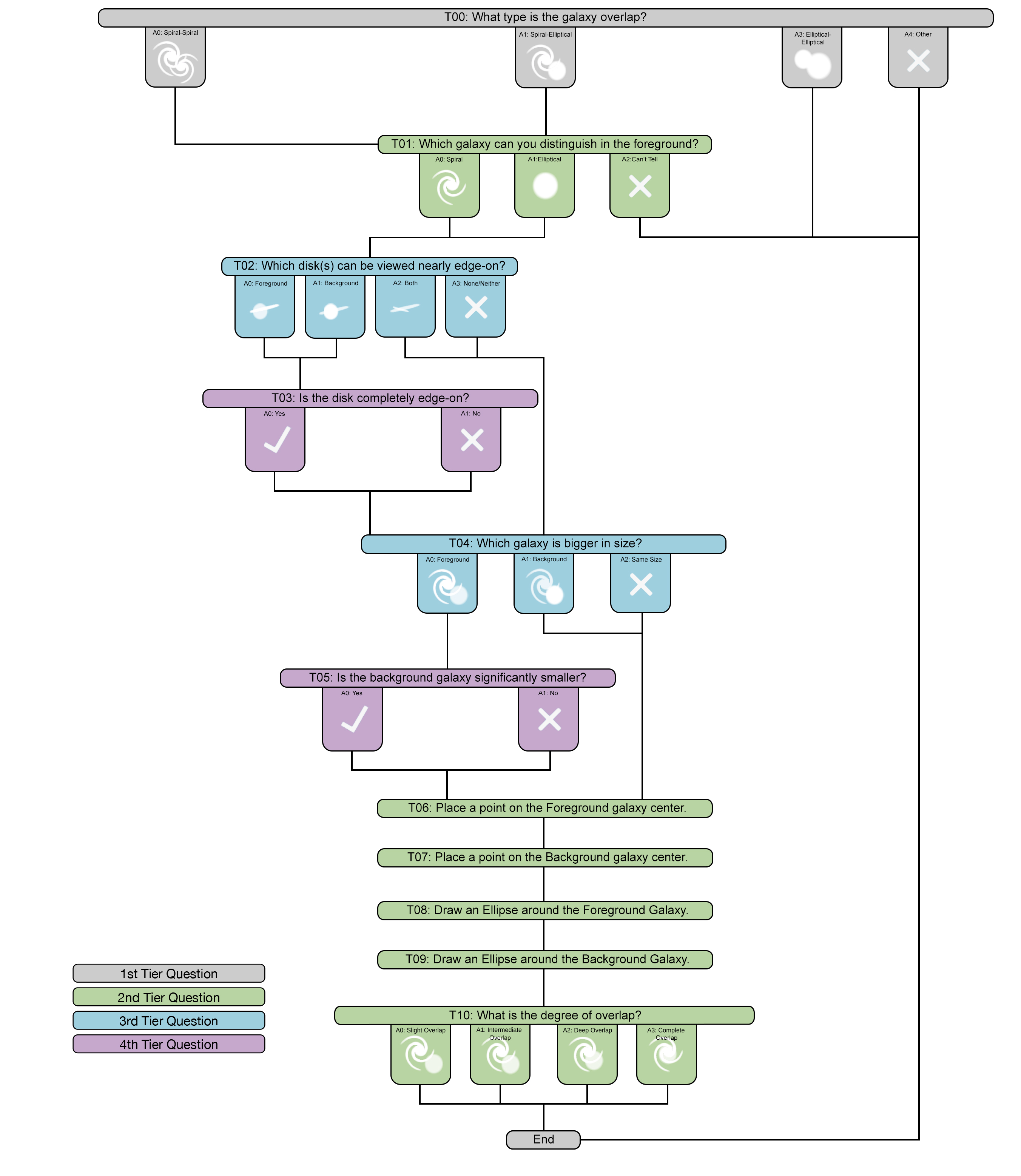}
\label{decision_tree}
\centering
\end{subfigure}%
\caption{Flowchart of the workflow presented to our internal classifiers. We labelled each question with its corresponding task number. Tasks are also colour-coded based on their tier level. Each tier, coloured grey, green, blue, and purple, represents zero, one, two, three, or four branching points in the decision tree, respectively.}
\label{DecisionTree}
\end{figure*}

For classification, we generated images of the candidate occulting pairs from \textit{legacystamps} \citep{Sweijen22}, a Python module that provides \textcolor{black}{JPEG} cutouts of DR10 from the DESI Legacy Survey. We downloaded each \textit{grz} image at a fixed native resolution of \(0.262\) arcsec per pixel. The images were then selectively rescaled to fit each occulting pair into the field of view (FOV) and subsequently resized to \(500\) x \(500\) pixels using Lanczos interpolation \citep{Lanczos1938} for use in \textit{Zooniverse}. The names of the image files included identifiers for the astronomical object (iauname) for use in cross-referencing, along with the \textcolor{black}{image} scaling factor (ISF), as our workflow involves pixel measurements (see T06-09; Figure \ref{DecisionTree}).

\subsection{Classification Workflow}
\label{workflow}


Since occulting pairs are distinctly different from standard non-occulting morphological classifications, it was essential to develop a new workflow, similar to traditional Galaxy Zoo projects, but tailored to classify occulting pairs and accommodate volunteer accessibility. To implement this workflow, we utilised the \textsc{Zooniverse Project Builder} software, which enables any user to easily create a simple citizen science project through a user-friendly browser interface. This platform allows volunteers to classify directly through their web-based interface, simplifying and streamlining the classification process. We note that due to the small-scale nature of this project (\(\sim\)2400 pairs), internally selected \textcolor{black}{student and public volunteers} completed the classifications through the website.

\renewcommand{\thefootnote}{\arabic{footnote}}
We designed the questions in our customised workflow (see Figure \ref{DecisionTree}) to be answered by a classifier with little astrophysical background while fitting the pair categories described by \cite{Keel13}. Given the citizen scientist nature and subject of this project, it was almost essential to draw inspiration from previous Galaxy Zoo projects and \cite{Keel13} 's work to create the most straightforward classification questions that yielded classifiers using the fewest possible questions. However, the actual effectiveness of our workflow remains uncertain. We intend to examine the accuracy of our approach in more detail in Section \ref{accuracy}.

As an additional aid to classifiers, we also provided a brief morphology tutorial to all users, which they were required to complete at the beginning of the workflow. Further assistance was also available on a per-task basis to support classifiers in answering these questions.

Our workflow for this project follows a hierarchical design similar to that of previous Galaxy Zoo projects. Every classification begins with the step of identifying whether the pair was a 'spiral-spiral,' a 'spiral-elliptical,' a 'elliptical-elliptical,' or a 'other' (artifact, star, multiple, or irregular) galaxy overlap. We excluded 'multiple' pairs due to their classification difficulty and 'irregular' pairs because of their lack of symmetry, which makes them difficult to use in dust studies. Likewise, we excluded 'elliptical-elliptical' overlaps from additional tasks, as it is challenging for individuals to decide on which elliptical is in the foreground or background from images lacking proper redshift measurements. Subsequent questions in the workflow depend on the previous answer(s) given by each classifier, with each classifier selecting a single answer before continuing to the next question. Classifiers also had the option to return to previous questions if needed. We note that we categorise the annotations created by tasks T06-09 as an answer.

After completing the workflow and classifying an image, we presented the user with a new image to classify. Each new image presented to classifiers was entirely random. Users continued to receive images until each image reached a retirement count of 4 classifications for each occulting pair. Although we initially set a higher limit, we decided to lower it due to the small-scale and beta nature of our citizen science project. For a larger, public project, we would opt for a higher number to ensure greater reliability of classifications. Later, we will also explore how this low classification count impacts the result. In total, the internal project ended with \(\sim\) 9,500 classifications from 15 registered and 47 unregistered participants for the 2,380 images selected to be classified. We note that the unregistered participants only contributed about \(4\%\) of the total classifications, while the registered participants classified the remaining \(96\%\).

\section{Data Reduction}
\label{data_reduc}

\subsection{Multiple Classifications}

In a small subset of cases, users submitted multiple classifications for a single object, either due to a technical issue with the website or deliberate reclassification. Regardless of the cause, to treat each vote as an independent measurement, these repeat classifications were excluded from the final dataset, leaving only the first classification for each object from each user. These repeat classifications occurred for fewer than \(1\%\) of pairs and did not have a significant impact on morphological classifications.

\subsection{Tied Classifications}

Given the low retirement count of 4 classifications per occulting pair, the inherent difficulty of classifying these pairs, and potential issues with our workflow (see Section \ref{limitations}), approximately \(23\%\) of our classifications resulted in a tie (up to \(56\%\) when excluding 'Other' classifications). To address this issue and ensure a valid morphological classification in the catalogue, we implemented a voting scheme based on majority rule and user consistency throughout the project.

\begin{figure}[!htbp] 
\centering
\begin{subfigure}{1\textwidth}
\graphicspath{ {Images/}}
\includegraphics[width=\textwidth]{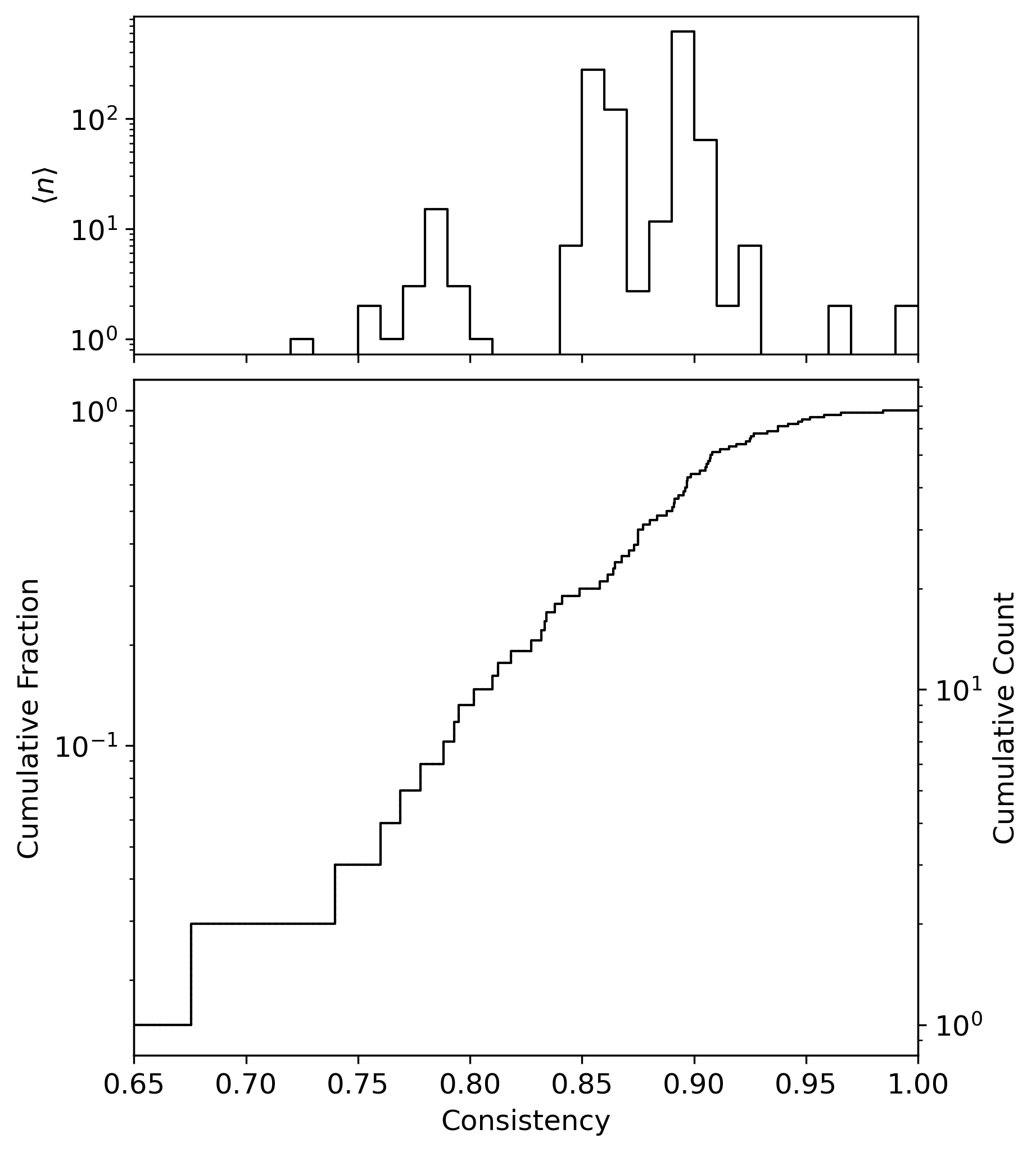}
\end{subfigure}%
\caption{The distribution of user consistency \(\kappa\) based on the methods outlined in \cite{Willett13, Willett17}. Bottom: the cumulative distribution of consistency for all users. Top: the average number of galaxy pairs, \(\langle{n}\rangle\), classified by users as a function of their consistency.}
\label{consistency}
\end{figure}

\subsubsection{User Consistency}
We calculated the consistency of each user using methods established in previous Galaxy Zoo projects \citep{Willett13, Willett17}. First, we aggregated individual user votes to calculate a vote fraction for each possible response \((f_r)\) to each task in the decision tree, excluding annotation tasks. We then compared each user's votes with these vote fractions to compute their consistency score, \(\kappa\):

\[\kappa = \frac{1}{N_r}\sum_{i=1}^{N_r}{\kappa_i}\]
\newline
where \(N_r\) is the total number of potential responses to a task and

\[
\kappa_i = 
\begin{cases}
f_r, & \text{if vote corresponds to this response,} \\
(1 - f_r), & \text{if vote does not correspond.}
\end{cases}
\]

Votes that agreed with the majority classification were assigned a high consistency value, while those that disagreed received a lower value. We assigned each user a mean consistency, \(\tilde{\kappa}\), by calculating the mean of the consistency values for each task. Figure \ref{consistency} shows the distribution of these values, together with the average number of galaxy pairs classified by each user.

\subsubsection{Tie Breaking}
\label{ties}

To resolve a tied classification \textcolor{black}{for a task}, we selected the classification provided by the user with the highest mean consistency as the definitive \textcolor{black}{classification} label. We also ensured that this user's classifications aligned with the majority before selecting their classification as the tiebreaker. This approach ensured that each occulting pair received a single, majority-based label while also leveraging user reliability to resolve uncertain pairs. Out of the 2,380 candidate pairs, 551 required tie correction \textcolor{black}{to one or more tasks}. Pairs with \textcolor{black}{task} classifications resolved through this method \textcolor{black}{were included in the catalogue} and are flagged in the morphological catalogue \textcolor{black}{(Section \ref{morphological})} as "\(t\)" to indicate that these labels do carry some additional uncertainty.

\section{The Catalogue}
\label{catalogue}

The Overlap Zoo Beta (hereafter OZb) catalogue includes three distinct sub-samples: vote counts and fractions for each task in the decision tree for each galaxy pair, pixel-based annotations for each galaxy pair, and morphological classifications based on the pair categories outlined in \cite{Keel13}, along with redshifts and colour magnitudes. Researchers can access the data for the three sub-samples from (insert link here). Included in this paper is a subset of each catalogue (Tables \ref{tab:user_class}, \ref{tab:user_anot}, and \ref{tab:occultpairs}).

\subsection{Classification Dataset}
Table \ref{tab:user_class} contains the classifications for the 2,380 candidate occulting galaxy pairs presented to volunteers via \textsc{Zooniverse}. We identify each galaxy pair by its unique J2000.00 name (iauname). The total number of classifications and the total number of votes for each response are denoted by \(N_{class}\) and \(N_{votes}\), respectively, following the terminology outlined in \cite{Willett13}. For each galaxy pair, we provide two additional parameters: the raw number of votes (\textbf{\(^\star\_count\)}) and the fraction of votes for said response (\textbf{\(^\star\_fraction\)}). Morphological classifications for this OZb subset are excluded from this table to maintain a consistent structure with the \cite{Keel13} catalogue and to remove pairs that are not genuine occulting candidates from the final morphological sample.

\begin{table*}[]
\centering
\caption{\textsc{Overlap Zoo Beta: User Classifications} }
\begin{adjustbox}{max width=\textwidth}
\begin{tabular}{ c c c 
| >{\centering\arraybackslash}p{2.8cm} >{\centering\arraybackslash}p{2.8cm} 
| >{\centering\arraybackslash}p{2.8cm} >{\centering\arraybackslash}p{2.8cm} 
| c }
\toprule
& & & 
\multicolumn{2}{c}{\textbf{t00\_overlap\_type\_a0\_spiral\_spiral\_}} & 
\multicolumn{2}{c}{\textbf{t00\_overlap\_type\_a1\_spiral\_elliptical\_}} &
\multicolumn{1}{c}{\textbf{...}} \\
Coordinate Name & $N_{\mathrm{class}}$ & $N_{\mathrm{votes}}$ & 
count & fraction & 
count & fraction & 
\\
\midrule
J000346.45+001153.2 & 4 & 18 & 0 & 0 & 2 & 0.5 & \\
J000415.41-015101.1 & 4 & 28 & 3 & 0.75 & 1 & 0.25 & \\
J000716.81-005559.1 & 4 & 17 & 0 & 0 & 1 & 0.25 & \\
J000750.43-004411.5 & 5 & 26 & 3 & 0.6 & 0 & 0 & \\
J000832.89+010220.1 & 4 & 19 & 1 & 0.25 & 0 & 0 & \\
J000836.26-003545.9 & 4 & 9 & 0 & 0 & 0 & 0 & \\
J001002.99-022404.4 & 4 & 26 & 0 & 0 & 2 & 0.5 & \\
J001039.11+002828.3 & 4 & 33 & 2 & 0.5 & 1 & 0.25 & \\
J001238.04+004949.4 & 4 & 9 & 0 & 0 & 0 & 0 & \\
J001703.68-025021.8 & 4 & 31 & 3 & 0.75 & 1 & 0.25 & \\
J001755.50+002003.9 & 5 & 31 & 5 & 1 & 0 & 0 & \\
J002024.55+004917.5 & 4 & 19 & 0 & 0 & 3 & 0.75 & \\
J002026.63-010339.3 & 4 & 18 & 0 & 0 & 2 & 0.5 & \\
J002134.87-005952.7 & 4 & 9 & 0 & 0 & 0 & 0 & \\
J002230.08-010438.9 & 4 & 9 & 0 & 0 & 0 & 0 & \\
J002319.00+003800.1 & 4 & 24 & 2 & 0.5 & 1 & 0.25 & \\
J002336.10+000547.3 & 4 & 9 & 0 & 0 & 0 & 0 & \\
J002338.28-004726.5 & 4 & 11 & 0 & 0 & 1 & 0.25 & \\
J002355.06-010540.1 & 4 & 9 & 0 & 0 & 0 & 0 & \\
\bottomrule
\end{tabular}
\end{adjustbox}

\label{tab:user_class}
\end{table*}

\begin{table*}[]
\centering
\caption{\textsc{Overlap Zoo Beta: User Annotations}}
\begin{adjustbox}{max width=\textwidth}
\begin{tabular}{ c c 
| >{\centering\arraybackslash}p{1.5cm} >{\centering\arraybackslash}p{1.5cm} 
| c
| >{\centering\arraybackslash}p{1.3cm} >{\centering\arraybackslash}p{1.3cm} >{\centering\arraybackslash}p{1.3cm} >{\centering\arraybackslash}p{1.3cm} >{\centering\arraybackslash}p{1.3cm} 
| c }
\toprule
& & 
\multicolumn{2}{c}{\textbf{t06\_foreground\_point\_center\_}} & 
\multicolumn{1}{c}{\textbf{...}} & 
\multicolumn{5}{c}{\textbf{t08\_foreground\_ellipse\_}} & 
\multicolumn{1}{c}{\textbf{...}} \\
Coordinate Name & isf & 
X & Y & 
& \(\theta\) (rad) & rX & rY & X & Y & 
\\
\midrule
J000346.45+001153.2 & 2.81 & 201.53 & 272.92 & ... & -1.52 & 21.71 & 19.69 & 201.31 & 271.56 & ... \\

J000415.41-015101.1 & 2.81 & 170.80 & 136.13 & ... & 0.29 & 111.79 & 114.25 & 170.36 & 134.63 & ... \\
J000415.41-015101.1 & 2.81 & 170.30 & 140.41 & ... & -2.42 & 92.40 & 76.51 & 169.64 & 133.74 & ... \\
J000415.41-015101.1 & 2.81 & 248.93 & 251.48 & ... & 1.27 & 165.51 & 82.75 & 249.54 & 250.91 & ... \\

J000716.81-005559.1 & 2.81 & 211.20 & 252.55 & ... & -1.57 & 11.57 & 10.23 & 211.06 & 251.89 & ... \\

J000750.43-004411.5 & 2.81 & 250.11 & 252.12 & ... & -2.39 & 116.03 & 58.02 & 247.71 & 254.53 & ... \\
J000750.43-004411.5 & 2.81 & 246.05 & 254.65 & ... & 0.76 & 101.09 & 42.03 & 230.31 & 289.66 & ... \\

J000832.89+010220.1 & 2.81 & 304.51 & 244.92 & ... & -0.71 & 122.00 & 61.00 & 295.71 & 240.13 & ... \\

J001002.99-022404.4 & 2.81 & 314.27 & 165.30 & ... & 1.31 & 52.99 & 26.50 & 314.44 & 164.56 & ... \\
J001002.99-022404.4 & 2.81 & 248.69 & 253.57 & ... & -0.07 & 217.27 & 71.55 & 249.44 & 252.10 & ... \\

J001039.11+002828.3 & 2.81 & 249.05 & 246.94 & ... & -4.55 & 130.70 & 27.21 & 249.00 & 249.88 & ... \\
J001039.11+002828.3 & 2.81 & 249.50 & 246.58 & ... & -1.43 & 125.33 & 29.95 & 248.98 & 246.43 & ... \\
J001039.11+002828.3 & 2.81 & 249.71 & 247.90 & ... & -4.55 & 138.54 & 35.64 & 248.91 & 252.70 & ... \\

J001703.68-025021.8 & 2.22 & 285.44 & 290.09 & ... & -1.85 & 59.87 & 15.22 & 284.34 & 289.31 & ... \\
J001703.68-025021.8 & 2.22 & 248.51 & 253.72 & ... & -1.24 & 97.15 & 48.57 & 240.51 & 244.12 & ... \\
J001703.68-025021.8 & 2.22 & 283.31 & 287.00 & ... & -1.88 & 66.07 & 23.09 & 289.31 & 290.00 & ... \\

J001755.50+002003.9 & 2.22 & 411.90 & 403.60 & ... & 0.02 & 55.87 & 58.00 & 412.07 & 402.97 & ... \\
J001755.50+002003.9 & 2.22 & 412.97 & 406.41 & ... & -0.23 & 54.67 & 43.82 & 410.30 & 399.74 & ... \\
J001755.50+002003.9 & 2.22 & 270.30 & 219.75 & ... & -1.21 & 214.72 & 135.97 & 267.50 & 216.95 & ... \\

\bottomrule
\end{tabular}
\end{adjustbox}
\label{tab:user_anot}
\end{table*}

\begin{table*}[!t]
\centering
\caption{\textsc{Overlap Zoo Beta: Morphological Classifications}}
\label{tab:occultpairs}
\begin{tabular}{c c c c c c c c c}
\hline
Coordinate Name & SDSS ObjID & DESI ObjID & $r_1$ & $r_2$ & $z_1$ & $z_2$ & Type $\Delta z$ & Cross-ID \\
\hline

J001755.50+002003.9 & 1237657191444050000  & -- & 14.89 & 16.96 &0.0180 & 0.0176 &S(I)  & MCG+00-01-059 \\

J003135.89+001001.6 & 1237663784200830000  & --  & 16.65  & 17.48  &  0.0728 & 0.0739 & Q(II)& -- \\ 

J004532.78+005657.5 & 1237663716556210000  & 39627809076810300 & 17.00 & 18.01 & 0.1104 & 0.1098  & Phi(III)  & -- \\

J005555.93+003940.2 &1237663204920260000  & 39627803078960800 & 15.67 & 16.73  &0.0671 & 0.0665 & Phi(II) & -- \\

J010436.00+000852.5 & 1237663784204430000  & 39627791037109200 & 17.04 & 17.59  & 0.0711& 0.1720 & X(I) $\Delta z$  & -- \\

J011535.02-005052.3 & 1237666338116200000  & --   & 14.58 & 12.84 & 0.0382 & 0.0058 & S(II) $\Delta z$ & NGC 450, UGC 807 \\

J013904.93+001748.1 & 1237657071157960000 &2842530165030910  & 16.15 & -- & 0.1237 & -- & E & -- \\ 

J014434.10+004418.7 &1237657071695420000  &1237657071695420000  & 15.08  & 16.73 & 0.0590 & 0.1007  & Q(II) $\Delta z$  & -- \\

J014448.57-000620.9 & 1237666407917020000  & 39627785165085100  & 16.33  & 18.37  & 0.0559 &0.1728  & Q(I) $\Delta z$  & -- \\

J021704.77+011439.0 &1237680099168150000  & 39627815502480100 &12.68  & 16.48 & 0.0214& 0.1223 &B(IV) $\Delta z$  & NGC 875 \\

J072535.99+250028.0 & -- & -- & -- & -- & -- & -- & F(I) & -- \\

J090049.81+043459.8 &1237658298986980000  & 2842634645143550 &14.87  & --  & 0.0946 & -- & B(IV) & LEDA 1270284 \\

J091032.80+040832.4 & 1237654606389700000 & 39627889674555300 & 18.08 & 15.82 & 0.0731 & 0.0733 & F(III) & -- \\

J094344.49+021442.7 & 1237653665257090000 & 39627841532334200 &16.93  &18.20  &0.0929 &0.0924  & Phi(II)  & -- \\

J100932.56-001414.5 &1237651800698250000  &--  &  15.35&--  &0.0697 & -- & B(IV) & -- \\

J102706.87+060333.2 &1237658423015570000  &--  &16.69  &--  &0.0589 &--  &X(II)  & LEDA 1293849 \\ 

J102756.99+055150.2 & 1237658298459610000 & 39627926131443500 & 16.97 & 15.87  & 0.1215 & 0.0561  & F(I) $\Delta z$ & --\\

J103545.36+200245.9 & 1237667782287220000 & -- & 14.62 & -- & 0.0551 & -- & SE(III) & MCG+03-27-064 \\

J104115.68+062141.0 & 1237658423017070000  & --  & 14.57 & 13.94 &0.0208 &0.0192  & SE(I) & UGC  5818 \\

J111016.87+020028.5 &1237651753468030000  & 39627835857438300 & 15.70 &17.87  &0.0754 &0.1413  &X(I) $\Delta z$  & LEDA  156322 \\

J112243.80-074033.9 & -- & -- & -- & --  & -- & -- & B(IV)  & MCG-01-29-015 \\

J112247.18-004122.4 &1237674649387720000  &39627769469997800  & 16.85  & -- &0.1450 & -- & Phi(II)  & 2dFGRS TGN373Z199 \\

J112948.29-044603.0 & --  & 39627672929702500 & -- &15.98  &-- & 0.0822 & S(I) & -- \\

J114503.87+195825.2 & 1237668293912820000 & 39628261617046700 & 12.89 & 15.10 & 0.0168& 0.0198 & B(II) & NGC  3861 \\

J114608.29-010709.7 & 1237674648853410000  & -- & 16.772 & -- & 0.1188 & --  &  E & -- \\

J120416.49+240847.9& 1237667910591640000 & -- &13.85  & 16.31 &0.0510 &0.0529  &SE(CT)  & UGC 7055a \\

J122614.44+001601.9 &1237648721770440000  & 39627793897620400 & 15.23 &18.73  & 0.0226 & 0.1051 & S(II) $\Delta z$ & Z  14-38 \\

J125345.43+200536.2 &1237667915960680000  & 39628261889673700 &15.53  & -- &0.0224 & -- & X(III)  & LEDA 1616133 \\

J130534.99+210803.0 & 1237667733976640000  & -- & 14.72  & 17.50 & 0.0619 & 0.0636  & E & IC 4170 \\

J131816.20+001939.7 & 1237651204231850000 & 39627794115724800 & 15.90 &17.68  & 0.0817&0.0819  & SE(II) & 2dFGRS TGN330Z132 \\

J133455.79+134428.9 &1237664289936370000  &--  & 12.73 &14.63  &0.0231 & 0.0227 & SE(III) & NGC  5222 \\

J140559.81+141411.6 & 1237662637973630000 & -- & 14.46 & 16.21 &0.0396 &0.0394  & E & LEDA   50288 \\

J140737.44-023536.9 & 1237655494372620000 & 39627727883472000 & 16.78 & 16.47 & 0.1385 & 0.1383 & F(I) & -- \\

J141652.80+104823.9 &1237661949730220000  & 39628046705101600 & 11.71 & 14.14 & 0.0247& 0.0229 & E & NGC  5532 \\

J143651.72+060824.1 &1237662267533420000  & --  & 16.07 & -- & 0.0588 &--  & S(I) & LEDA 1295240 \\

J144311.82+185247.8 & 1237667735597280000 & --  & 13.42 & --  & 0.0317 & -- & S(III) & NGC  5737 \\

J145116.70+184128.9 & 1237667783387180000 & -- & -- & 13.95 & -- & 0.0435 & F(III) & IC 1062 \\

J154543.89+112422.1 & 1237668332571850000  &--  &16.88  &--  &0.0500 &--  & Q(I)  & -- \\

J154954.44+085140.6 & 1237662533823430000 & 39627999431100400 & 15.25 &16.31  &0.0737 &0.0736  & F(II)& -- \\

J203759.02-004122.4 &1237656567573900000  &39627771797841300  &17.81  &19.17  &0.1018 & 0.1011  & Q(I)  & -- \\

J212622.50+020440.0 & -- & 39627838441130800 &14.39 -- &  & 0.04855 & -- & B(I) & UGC 11742 \\

J213231.86+042355.6 & 1237669761180950000 &39627898805557000  &16.91  & 16.86  &0.0790 &0.0818  & Phi(I) & -- \\

J215556.94-065338.0 & 1237680241436910000 & -- &15.05  & -- &0.0658 &--  & S(II) & -- \\

J223124.73+004646.7 & 1237663544223130000 & 39627808514772300 &16.90  &18.59  & 0.1322& 0.1311 & SE(IV) & LEDA 1175253 \\

J223614.90+012444.9 & -- & 39627826655135800 &--  & 16.32 & --& 0.0625 & S(I) & LEDA  193199 \\

J232245.93-000641.1 &1237657190901150000  &39627790609291700  &17.59  &19.97  & 0.1219& 0.1218 &X(III)  & -- \\

J233404.08+013507.0 &1237678595940150000  & --   & 16.42 &  --& 0.1331 & -- & S(III) & VV 187a \\

J235734.51+020455.2 & 1237678618491280000  & -- &16.02  & -- &0.0760 & -- & SE(CT)  & LEDA 1215402 \\

\hline
\end{tabular}%
\end{table*}

\subsection{Annotation Dataset}
\textcolor{black}{Table \ref{tab:user_anot} lists the 1,428 pixel-based annotations (including duplicates), derived from annotation tasks T06–T09 in Figure \ref{DecisionTree}.} Each galaxy in this subset is, again, identified by its unique J2000.0 name. The corresponding ISF value is also included, as it is necessary for accurate use of the pixel measurements (see Section \ref{images}). Point tool annotation values for the foreground and background galaxy (Task T06-T07) are denoted by \(X\) (\textbf{\(^\star\_X\)}) and \(Y\) (\textbf{\(^\star\_Y\)}), \textcolor{black}{with the associated right ascension (RA) and declination (DEC) for each measurement also included.} Ellipse tool annotation values for the foreground and background galaxy (Task T08-T09) are denoted by \(X\) (\textbf{\(^\star\_X\)}), \(Y\) (\textbf{\(^\star\_Y\)}), \(rX\) (\textbf{\(^\star\_rX\)}), \(rY\) (\textbf{\(^\star\_rY\)}), and \(\theta\) (\textbf{\(^\star\_\theta\)}). In both cases, \(X\) and \(Y\) represent the centre coordinates of the tool, while \(rX\), \(rY\), and \(\theta\) represent the horizontal radius, vertical radius, and orientation angle in radians of the ellipse, respectively.

By including these annotations in the OZb catalogue, we hope to expand the application of various analysis techniques across a larger sample of candidate occulting pairs. The annotations in tasks T06-T09 are not only necessary for querying (see Section \ref{Query}), but also for applying various analysis techniques in and based on \cite{White92}'s work. This includes differential photometry techniques such as rotational subtraction and isophotal fitting and subtraction. Both approaches enable accurate estimation of key nuisance parameters, including dust extinction and attenuation. The latter differential approach was recently applied with great success to the occulting pair VV191a/b \citep{Keel23, Robertson24}. Our goal is to extend these techniques to more pairs in order to increase the statistical variance and robustness of the results of the current pairs, such as the remarked pair VV191a/b.

\subsection{Morphological Dataset}
\label{morphological}
Table \ref{tab:occultpairs} contains the morphological classifications for \textcolor{black}{765} occulting galaxy pairs\textcolor{black}{, out of the initial 2,380,} that could be categorised based on the pair categories outlined in \cite{Keel13}. \textcolor{black}{The remaining 1,615 pairs that could not be categorised were those classified as "Other" in Task T00 (Figure \ref{DecisionTree}), and are excluded in this subset of the catalogue.} We identify each occulting pair by its unique J2000.0 coordinates, SDSS ID, and DESI ID. Using our new querying methodology (see Section \ref{Query}), we obtain magnitude and spectroscopic redshift measurements from SDSS \citep{SDSS-DR18} and DESI \citep{DESI-DR1}, denoting the central galaxy's values as \textit{r1} and \textit{z1}, and the off-centre galaxy values as \textit{r2} and \textit{z2}. While DESI does not directly provide magnitude measurements, we calculated the r-band magnitudes from the provided fluxes in nanomaggies using the standard zero-point conversion. However, we warrant caution when interpreting these magnitudes, as blending from overlapping galaxies may impact their accuracy \citep{Keel13}. The subset included in this paper comprises all systems shown in Figure \ref{classification_images}, along with a few additional interesting pairs, listed in order of increasing RA.

\subsubsection{Pair Categories}
\label{paircat}
Below, we briefly summarise each mnemonic designation from the categories defined in \citet{Keel13}, along with one additional category of our own. For each designation, we outline the specific tasks and corresponding responses used for classification. We also provide a summary of the number of pairs in each designation in Table \ref{T0_Classifications}.

\begin{enumerate}
    \item \textbf{F:} a nearly face-on spiral in front of a smooth elliptical or S0 background system [Task T0 (A1), Task T01 (A0), and Task T02 (A3)].
    \item \textbf{Q:} a background galaxy nearly edge-on and projected nearly radially to the foreground galaxy [Task T0 (A0, A1), Task T01 (A0, A1), Task T02 (A1), Task T08, and Task T09].
    \item \textbf{\(\Phi\):} an essentially edge-on spiral backlit by a smooth elliptical or S0 background system [Task T0 (A1), Task T01 (A0), Task T02 (A0), and Task T03 (A0)].
    \item \textbf{X:} both galaxies are nearly edge-on with their disks essentially crossing [Task T0 (A0), Task T01 (A0), and Task T02 (A2)].
    \item \textbf{SE:} spiral/elliptical superpositions that do not fall in one of the other categories [Task T0 (A1), Task T01 (A1), and Task T02 (A0, A2, A3)].
    \item \textbf{S:} spiral/spiral overlaps. Useful for probing extinction and attenuation in the UV [Task T0 (A0), Task T01 (A0), and Task T02 (A0, A3)].
    \item \textbf{B:} the background galaxy has a much smaller angular size than the foreground galaxy [Task T0 (A0, A1), Task T04 (A0), and Task T05 (A0)].
    \item \textbf{E:} elliptical or S0 pairs [Task T0 (A3)].
    \item \textbf{\(\Delta{z}\):} pairs with a known redshift distance large enough to rule out interaction. Pairs with \(\Delta{z} > 0.008\) fall under this category.    
    \item \textbf{CT:} pairs that volunteers could not confidently identify the foreground/background distinction [Task T0 (A0, A1) and Task T01 (A2)].
    
\end{enumerate}

\subsubsection{Overlap Types}
In addition to these categories, we introduce further designations to describe the degree of overlap between the two galaxies in each pair, based on Task T10. These are defined as follows: 

\begin{enumerate}
    \item \textbf{I}: slightly overlapping galaxies. [Task T10 (A0)].
    \item  \textbf{II}: intermediately overlapping galaxies [Task T10 (A1)].
    \item \textbf{III}: deeply overlapping galaxies [Task T10 (A2)].
    \item \textbf{IV}: completely overlapping galaxies [Task T10 (A3)].
\end{enumerate}

Note that we assume all E-type pairs are complete overlaps, as we rarely observe significant separation between two elliptical galaxies. Again, we provide a summary of the counts for each overlap type in Table \ref{T10_Classifications}.

\begin{table}[ht]
    
    \centering
    \begin{tabular}{clc}
        \hline
        Pair Type & \multicolumn{1}{c} {Description} & Number \\
        \hline
        F  & Face-on spiral and background E/S0.                 & \textcolor{black}{42}  \\
        Q  & Background galaxy edge-on and radial.               & \textcolor{black}{35}  \\
        $\Phi$ & Foreground disk edge-on and background E/S0.    & \textcolor{black}{48}  \\
        X  & Crossing edge-on disks.                            & \textcolor{black}{39}  \\
        SE & Spiral/Elliptical pairs not otherwise listed.       & \textcolor{black}{125} \\
        S  & Two spirals.                                        & \textcolor{black}{110} \\
        B  & Background galaxy has a small angular size.         & \textcolor{black}{24}  \\
        E  & Two E/S0 galaxies.                                  & \textcolor{black}{342} \\
        $\Delta z$ & Redshifts indicate not physically associated. & 131 \\
        CT & F/B distinction is ambiguous. & \textcolor{black}{89} \\
        \hline
    \end{tabular}
    \caption{A summary of OZb classifications based on the pair types from \cite{Keel13}.}
    \label{T0_Classifications}
\end{table}

\begin{table}[!ht]
    \centering
    \begin{tabular}{clc}
        \hline
        Overlap Type & \multicolumn{1}{c} {Description} & Number \\
        \hline
        I  & Slight Overlap & \textcolor{black}{129}  \\
        II & Intermediate Overlap & \textcolor{black}{110}  \\
        III & Deep Overlap    & \textcolor{black}{62}  \\
        IV  & Complete Overlap  & \textcolor{black}{375}  \\
        \hline
    \end{tabular}
    \caption{A summary of the OZb classifications based on the degree of overlap between galaxy pairs.}
    \label{T10_Classifications}
\end{table}

\begin{figure*}[!htbp] 
\centering
\begin{subfigure}{1.0\textwidth}
\graphicspath{ {Images/}}
\centering
\includegraphics[width=\textwidth]{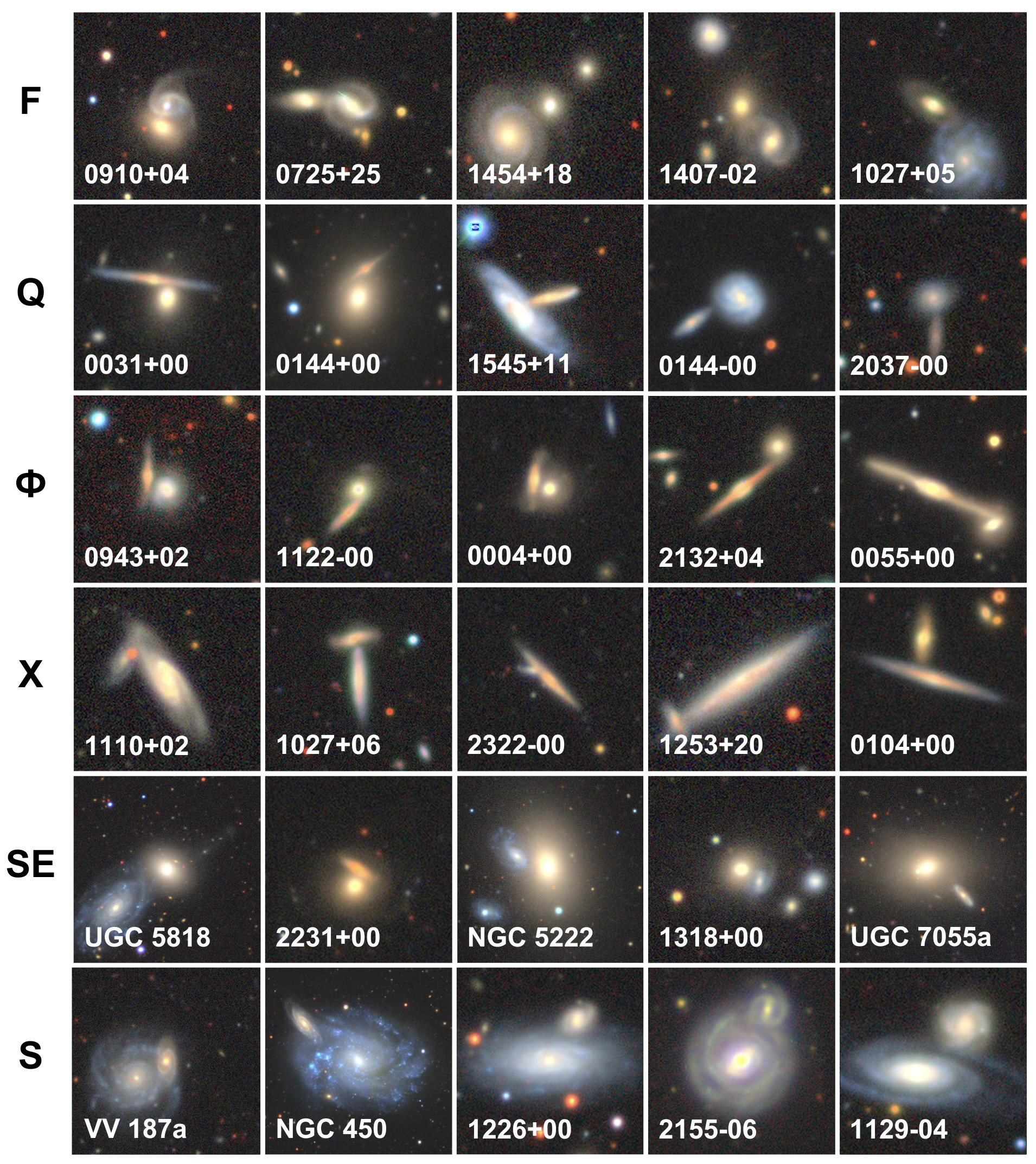}
\centering
\end{subfigure}%
\caption{Sample pairs based on the pair types described in \cite{Keel13}. PNG images were obtained from the Legacy Survey Viewer using the \textit{legacystamps} Python module \citep{Sweijen22} and independently verified for use in this figure. We identify pairs by their common name (e.g., NGC 450) or truncated coordinates (e.g., 1226+00).}
\label{classification_images}
\end{figure*}

\begin{figure*}[!htbp]
    \ContinuedFloat
    \centering
    \begin{subfigure}{\textwidth}
    \graphicspath{ {Images/}}
    \centering
    \includegraphics[width=\textwidth]{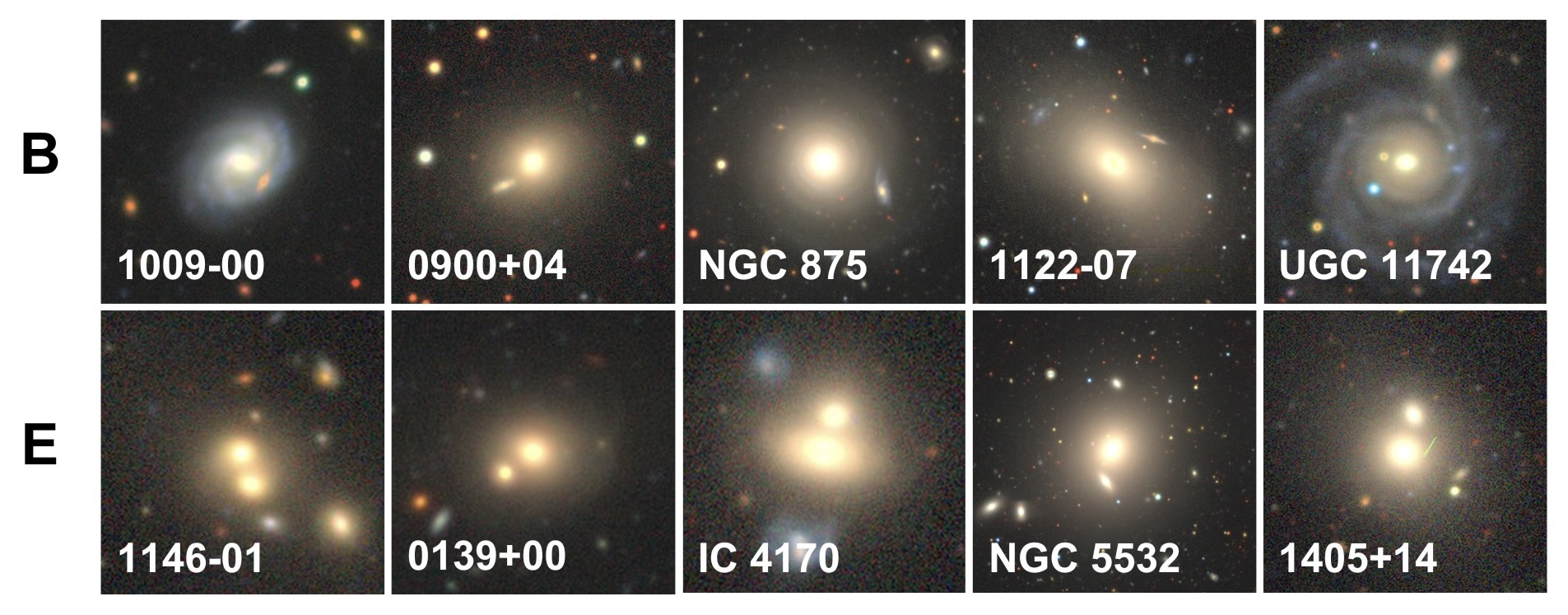}
    \centering
    \end{subfigure}%
    \caption[]{The remaining sample pairs following \cite{Keel13} mnemonics.}

\end{figure*}

\subsubsection{Querying}
\label{Query}
To obtain the measurements for the catalogue, we developed a new querying method tailored to occulting galaxy pairs. Unlike single, isolated galaxies in the Galaxy Zoo project, overlapping pairs require two distinct coordinate inputs and thus require manual annotation of the off-centre companion.

Since tasks T06–T07 (Figure \ref{DecisionTree}) address only foreground and background classification, we first had to determine which galaxy occupied the central and off-centre position using these annotations. To do this, we converted the annotated pixel coordinates into RA and DEC to estimate the centres of both galaxies, then compared these to the provided J2000.0 coordinates. We designated the galaxy with coordinates closest to the J2000.0 reference as the central galaxy, and the other as the off-centre companion. Using these coordinates, we then queried and assigned measurements such as object IDs, spectroscopic redshifts, and r-band magnitudes from both SDSS \citep{SDSS-DR18} and DESI \citep{DESI-DR1} for the central (\textit{z1} and \textit{r1}) and off-centre companion (\textit{z2} and \textit{r2}), prioritizing the DESI measurement when valid. To ensure that these queries returned only the measurement corresponding to the intended galaxy centre, we applied a strict separation tolerance of 0.00105 degrees (\(\sim 3.6\) arcseconds) from the input coordinates.\textcolor{black}{This value is much larger than the astrometric precision of DR10 (\(\sim 0.2\) arcseconds; \citealp{Dey19}), so even in cases of minor coordinate uncertainties, the correct galaxy measurement should be reliably selected.} For cross-identifications, we employed a similar query approach using SIMBAD \citep{SIMBAD}, albeit with less restrictive criteria relying solely on the central galaxy's position.

Manual inspection of the queried subsample showed that our querying methodology was \(98-99\%\) effective, with only a few problematic pairs. Most of the issues stemmed from proximity that fell under our strict tolerance, overcrowded fields, or occasional query failures. In such cases, we manually cross-referenced the DESI Legacy Survey with SDSS and DESI data to assign the correct IDs and redshifts. Although this level of accuracy was sufficient for the current project, we hope to improve it further in future iterations if possible.

Of the 765 galaxies queried, \textcolor{black}{22} centre and \textcolor{black}{403} off-centre pairs were also missing spectroscopic measurements necessary for accurate foreground, background distinction. We initially considered substituting missing values with photometric redshifts. However, these estimates are often unreliable. As shown in \citet{Keel13}, photometric redshifts can have errors exceeding \(5\sigma\) traceable to deblending problems. We chose to forgo them for these reasons.

\section{Classification Accuracy}
\label{accuracy}
To assess the accuracy of the classifications in the OZb catalogue, we currently employ three methods: (1) performing internal \textcolor{black}{diagnostic} tests comparing redshift difference \((\Delta{z} = |z_f -z_b|)\) and magnitudes (\(r_f\) and \(r_b\)) against the \textcolor{black}{fraction of classifications where the visual classifier correctly (\(z_f < z_b\)) or incorrectly (\(z_f > z_b\)) identified the foreground galaxy based on annotation positioning}, (2) comparing non-expert morphological classifications to those performed by an expert \citep{Keel13}, \textcolor{black}{and (3) comparing morphological classifications with GZD for S-, X-, and E-type pairs}. These approaches allow us to assess how accurately people can distinguish foreground from background galaxies, which can be difficult to visually distinguish depending on image quality and redshift separation, as well as how reliably non-experts and experts classify occulting pairs according to the morphological categories described in Section \ref{morphological}.

\subsection{Overlap Zoo Beta}

\subsubsection{Redshift Separation}
\label{redshiftsep}
\begin{figure}[!htb] 
\centering
\begin{subfigure}{1\textwidth}
\graphicspath{ {Images/}}
\centering
\includegraphics[width=\textwidth]{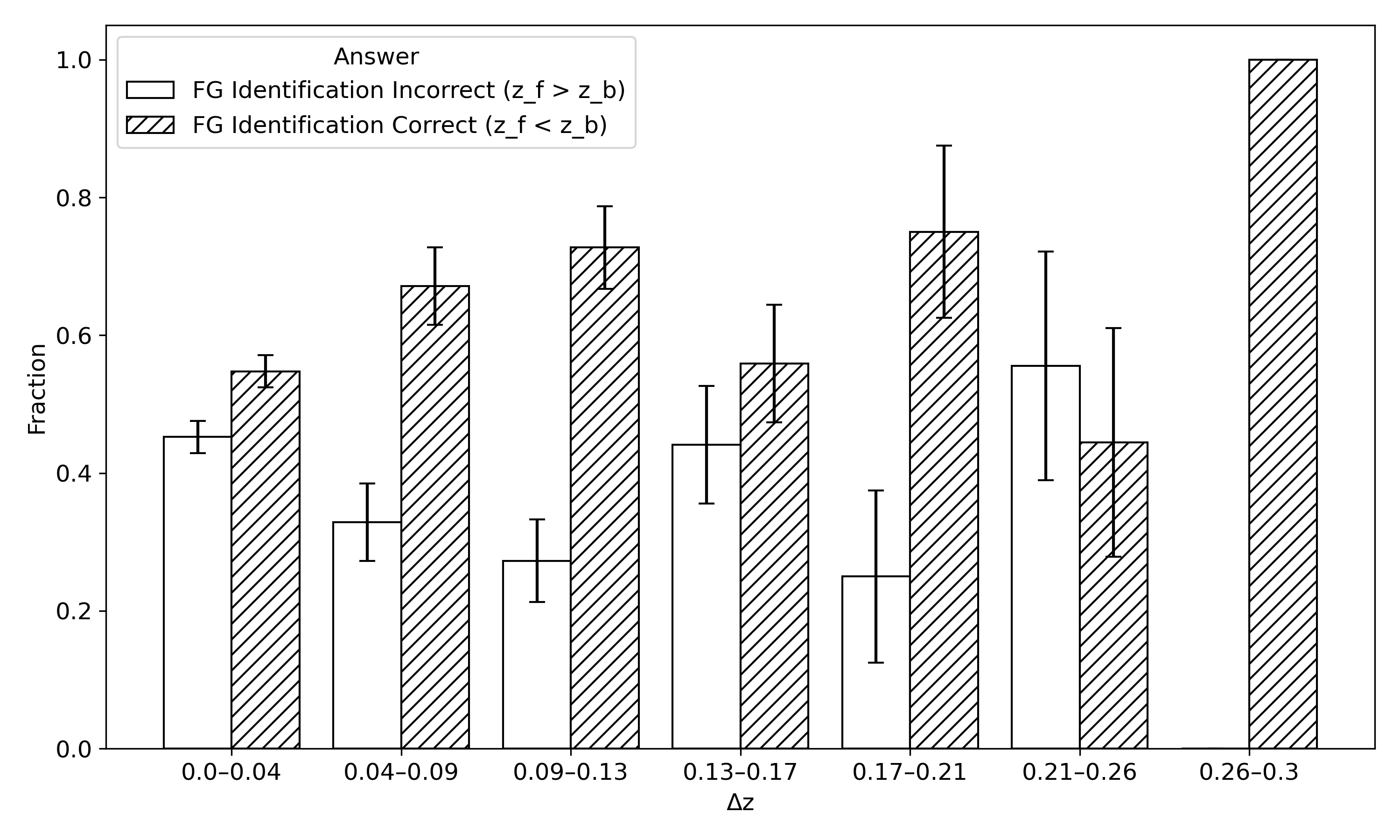}
\centering
\end{subfigure}%
\caption{The fraction of galaxy pairs \textcolor{black}{with foreground identification} correctly and incorrectly classified by users as a function of \(\Delta z\). We include only pairs with valid pixel-based redshift measurements in the sample.}
\label{redshift_accuracy}
\end{figure}

Figure \ref{redshift_accuracy} shows the fraction of occulting galaxy pairs \textcolor{black}{for which the visual classifier correctly and incorrectly annotated the foreground galaxy}, plotted against the redshift difference \(\Delta z\). We restrict the sample to pairs with valid pixel-based annotations, as our querying methodology required them (see Section \ref{Query}), and to pairs with two valid spectroscopic redshifts needed to calculate the redshift difference.

Since GZD includes a majority of galaxies under \(z=0.2\), the bulk of our redshift data will and does lie in the low \(\Delta{z}\) range (0.0-0.013), decreasing in sample size as the \(\Delta{z}\) increases (451 pairs in the 0.0–0.04 bin, 70 in 0.04–0.09, and 55 in 0.09–0.13). Within this range, we find that the ability to identify the foreground galaxy correctly improves as the redshift difference between pairs increases. However, at the lowest redshift separation, classifiers identify the foreground galaxy correctly nearly as often as they do incorrectly. Although higher-quality images could potentially make this distinction easier, visual classifications with minimal separation are likely unreliable at the current image resolution and depth. Beyond this range, our sample size decreases drastically, and we no longer observe this pattern. Large separations would likely make the distinction easier, though a bigger sample size at higher separations is necessary to confirm this.

\subsubsection{Magnitudes}
\label{magnitude}

\begin{figure}[!htb] 
\centering
\begin{subfigure}{1\textwidth}
\graphicspath{ {Images/}}
\centering
\includegraphics[width=\textwidth]{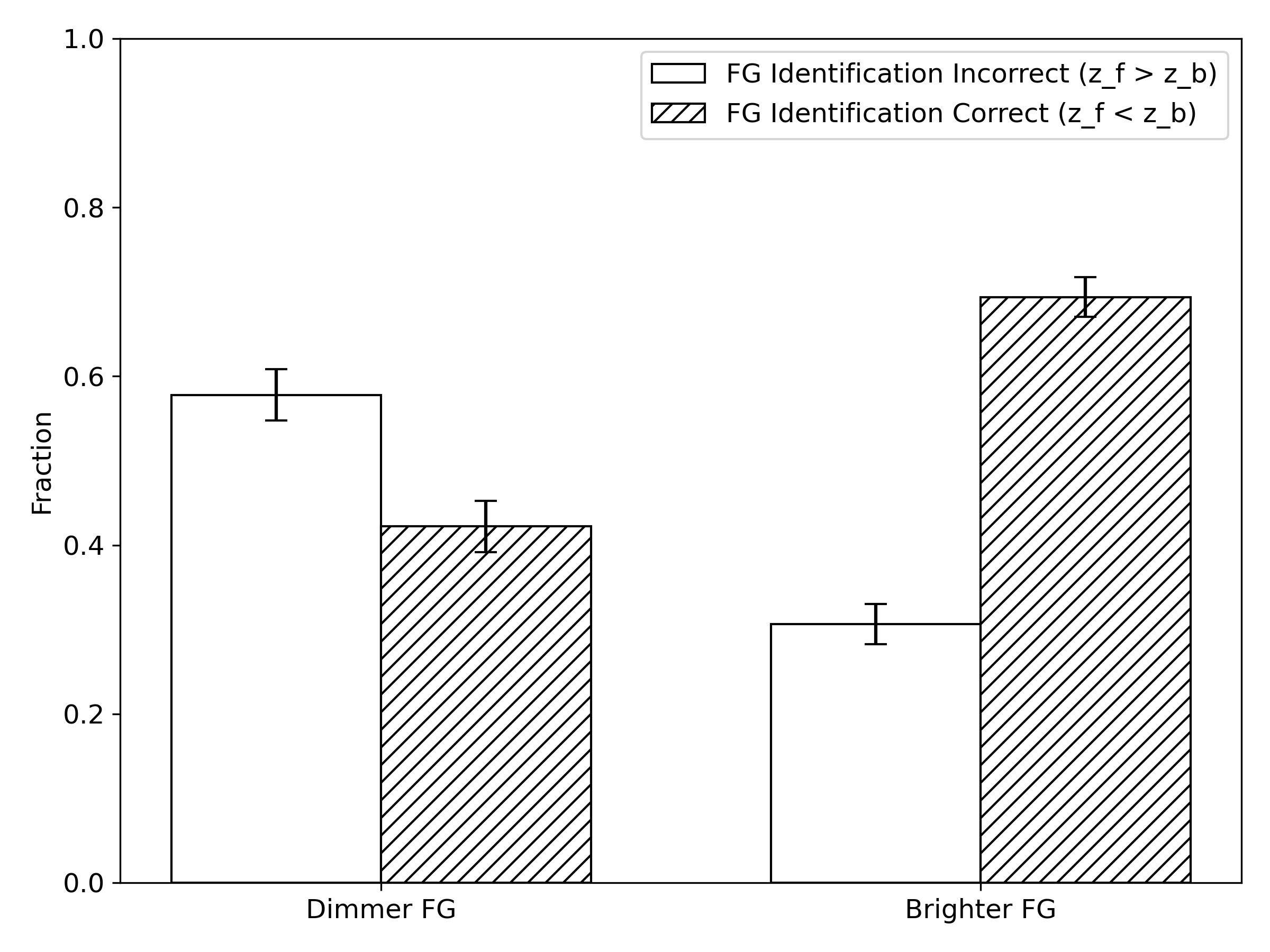}
\centering
\end{subfigure}%
\caption{The fraction of galaxy pairs \textcolor{black}{with foreground identification} correctly and incorrectly classified, based on whether the foreground galaxy appeared dimmer (\(r_f > r_b\)) or brighter (\(r_f < r_b\)) in \(r\)-band magnitude. We include only pairs with valid pixel-based redshift measurements in the sample.
}
\label{magnitude_accuracy}
\end{figure}

Figure \ref{magnitude_accuracy} shows the fraction of occulting galaxy pairs \textcolor{black}{for which the visual classifier correctly and incorrectly annotated the foreground galaxy}, based on whether the foreground galaxy was brighter (\(r_f < r_b\)) or dimmer (\(r_f > r_b\)) in r-band magnitude. Again, we restricted the sample to pairs with valid pixel-based annotations that had two valid spectroscopic measurements.

Unsurprisingly, pairs with brighter foreground galaxies have their positioning identified correctly more often than those with a dimmer foreground galaxy \textcolor{black}{(\(31\% \text{ vs } 69\%\))}. Surprisingly, however, pairs with a dimmer foreground galaxy show nearly comparable rates of correct and incorrect identification but are more often incorrectly identified \textcolor{black}{(\(58\% \text{ vs } 42\%\))}. One possible explanation is that when the foreground galaxy is fainter, its features may blend with those of the background galaxy, making the distinction between the foreground and background galaxies more difficult. In contrast, a brighter foreground galaxy, especially in Spiral–Elliptical systems, likely shows more distinct features and suffers less from blending, aiding the user in distinguishing foreground from background. However, these predicted magnitudes may be affected by the aforementioned blending effects (see the beginning of Section \ref{morphological}) and could introduce some uncertainty into our comparisons.

\begin{figure*}[!htbp]
    \centering
    \begin{subfigure}[]{0.48\linewidth}
        \centering
        \includegraphics[width=\linewidth]{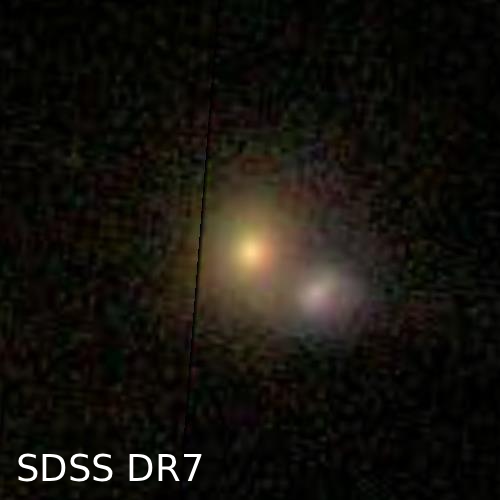}
        \label{fig:imageA}
    \end{subfigure}
    \hfill
    \begin{subfigure}[]{0.48\linewidth}
        \centering
        \includegraphics[width=\linewidth]{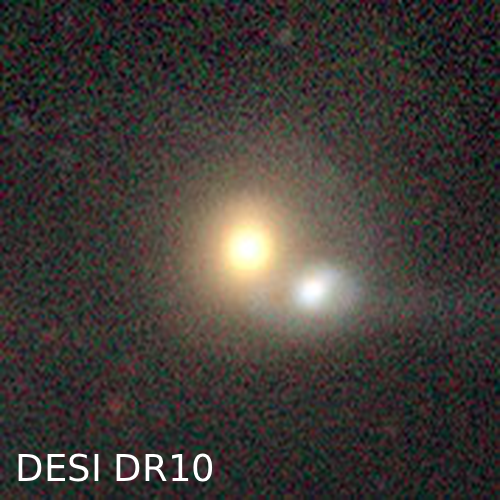}
        \label{fig:imageB}
    \end{subfigure}
    \caption{A comparison of image quality between SDSS and DESI using the galaxy pair \textit{J095303.71+064237.6}. \textit{Left}: The SDSS DR7 image used for classification in \cite{Keel13}. \textit{Right}: The DESI Legacy Survey DR10 image used for classification in this work. Both images match in size and pixel scale for direct comparison.}
    \label{comparison}
\end{figure*}

\subsection{OZb vs. Expert Visual Classification}
\label{extcata}


\textcolor{black}{To verify the effectiveness of} the citizen science approach \textcolor{black}{in classifying occulting pairs}, we compared our user-classified pairs with those assigned by an expert \citep{Keel13}. Of the \textcolor{black}{765} galaxy pairs in our catalogue, 189 overlap with the expert sample. Among these, \textcolor{black}{49} morphological classifications \textcolor{black}{(\(26\%\))} matched between our catalogue and the expert's. While this relatively low level of agreement may initially suggest an issue, several plausible explanations exist for the mismatch and other discrepancies.

A key factor likely influencing the discrepancy is the difference in imaging depth between SDSS DR7 \citep{SDSS-DR7} and the DESI Legacy Survey DR10 \citep{Dey19}. As a more recent survey equipped with improved instrumentation, DESI delivers images with greater depth (\(1.5-2\) magnitudes deeper) than SDSS. Figure \ref{comparison} presents a direct comparison of the same galaxy pair, as imaged by both surveys and classified by both the expert and this work. The difference in image quality between the two surveys is substantial enough that structural features of the background galaxy, clearly resolved in DESI, are almost indistinguishable in SDSS. \textcolor{black}{This has direct implications for morphological classification; in cases where the background galaxy's structure may be subtle or faint, classifiers using SDSS DR7, regardless of expertise, would likely be unable to identify key cues necessary to assign a distinct classification. Indeed, the classifications of both works reflect this: the pair in this work was classified as "Other", as there are visible signs of merging in the DESI image, whereas the expert classified it as an F-type (see Section \ref{paircat}), as there are no distinguishable merging signs due to the lack of visible structure in the SDSS image.}

Although not available for every pair, the expert catalogue also had access to spectroscopic and/or photometric redshift information to aid in distinguishing the foreground and background galaxies. In contrast, our classification process did not present redshifts directly to users, which could also explain the low agreement \textcolor{black}{in pair types where positional orientation matters}. Manually providing these measurements becomes impractical as our datasets scale; hence, we did not offer them directly (also see Section \ref{Query}). One of our primary goals was to assess non-experts' ability to distinguish foreground and background galaxies. Including redshifts would undermine this goal.

\subsection{OZb vs. GZD-1, -2, and -5}
\label{ozbvsgzd}

\textcolor{black}{While less robust than a direct comparison with an occulting pair catalogue, we also evaluate the accuracy of our classifications using single-galaxy morphological classifications from GZD-1, -2, and -5. However, the current iteration of our catalogue does not yet allow for a comprehensive comparative analysis across all morphological categories (see Section \ref{workflow}). Despite these limitations, comparisons with S-, X-, and E-type occulting pair morphologies remain feasible, as positional morphological knowledge of the central galaxy is not required for these types. For these comparisons, we use raw and debiased vote fractions from tasks T00 (A0 and A1) and T01 (A0) of the GZD decision tree (see Figure 4; \citealp{Walmsley22}), adopting an agreement threshold of 0.5; pairs with vote fractions exceeding this threshold are considered to have an agreeable classification.}

\textcolor{black}{Table \ref{tab:morphology_comparison} presents the agreement metrics based on the raw vote fractions for morphological answers in GZD, which correspond directly to the aforementioned morphological types of occulting pairs in our catalogue. Any agreement or disagreement in this comparison reflects a more direct view of how human observers perceive galaxy morphology from the images they are shown. Based on these comparisons, we find that, despite our relatively low classification count of 4 votes per pair, the classifications provided in our catalogue are relatively in agreement with the higher number of visual classifications in the GZD datasets. However, agreement could be improved, especially for X-types, as they show the lowest agreement between our datasets. The limited number of votes could be contributing to this lower agreement, as could the image quality (see Section \ref{imagegen}), or it may be due to redshift bias, as galaxies at higher redshifts are much harder to distinguish by the key features needed to identify these types. }

\textcolor{black}{To test whether this redshift bias is present, we present the agreement metrics based on the de-biased vote fractions in Table \ref{tab:morphology_comparison_debiased}, following the same structure as Table \ref{tab:morphology_comparison}. In GZD, redshift de-biasing mitigates classification bias by estimating how each galaxy would appear if observed at a fixed low redshift.. This reference redshift is set to 
\(z=0.02\) \citep{Walmsley22}, with the full de-biasing method described in detail by \citet{Hart16}.}

\begin{table}[ht]
\centering
\begin{tabular}{ccccccc}
\hline
 & \multicolumn{2}{c}{S} & \multicolumn{2}{c}{X} & \multicolumn{2}{c}{E} \\
  &  $N$ & \% & $N$ & \% & $N$ & \% \\
\hline
\multicolumn{7}{c}{GZD-1/2} \\
'smooth'  & -- & -- & -- & -- & 113 & 72.5 \\
'featured-or-disk' & 44 & 81.2 & 17 & 58.8 & -- & -- \\
'edge-on'  & -- & -- & 17 & 82.3 & -- & -- \\
\hline
\multicolumn{7}{c}{GZD-5} \\
'smooth'  & -- & -- & -- & -- & 290 & 83.4 \\
'featured-or-disk' & 88 & 85.2 & 28 & 71.4 & -- & -- \\
'edge-on'  & -- & -- & 28 & 67.8 & -- & -- \\
\hline
\end{tabular}
\caption{This table compares the agreement between the raw vote fractions for morphological types in GZD and the corresponding OZb types. A pair is considered agreeable if it has a likelihood of at least 0.5 (i.e., a majority) for the corresponding morphological type. Here, \textit{N} denotes the total number of overlapping pairs with GZD, while \textit{\%} indicates the fraction of those pairs that are in agreement.}
\label{tab:morphology_comparison}
\end{table}

\textcolor{black}{Comparing the two, we find that the agreement percentage for E-types is strongly affected, showing an approximately eightfold difference in GZD-1 and -2, and a twofold difference in GZD-5. This is consistent with the findings of \citet{Walmsley22}, who reported that the de-biasing procedure tends to over-correct the “Smooth or Featured” vote fractions, effectively reversing the original observed redshift trend (see their Figure 10). For X-type pairs, agreement follows the same reversing pattern and improves slightly for the “featured or disk” category, while the “edge-on” agreement exhibits a different trend. \citet{Walmsley22} did not observe a comparable change in “edge-on” vote fractions with redshift de-biasing, and it is likely that a marginal fraction of our edge-on pairs may be misclassified in our sample. On the other hand, we do not observe a comparable effect for S-type pairs as seen in E-types and X-types. We observe only a minor change in agreement compared with GZD-5, suggesting that most S-types in our sample are likely readily identifiable, large, luminous systems with well-defined disks or features, that are not greatly affected by redshift bias.}

\textcolor{black}{It is important to note, however, that the redshift de-biasing procedure itself carries significant uncertainties. The corrections applied rely on assumptions and models about how a galaxy would appear at a fixed low redshift. As a result, the agreement metrics from the de-biased sample may be highly uncertain, and these comparisons with the de-biased data should be interpreted cautiously. Some of the trends we observe, particularly in E- and X-type pairs, may be amplified by these uncertainties and by our low sample size, meaning that the true underlying agreement may differ significantly from that suggested by the de-biased vote fraction comparison. The true agreement is likely between our results here.}

\begin{table}[ht]
\centering
\begin{tabular}{ccccccc}
\hline
 & \multicolumn{2}{c}{S} & \multicolumn{2}{c}{X} & \multicolumn{2}{c}{E} \\
  &  $N$ & \% & $N$ & \% & $N$ & \% \\
\hline
\multicolumn{7}{c}{GZD-1/2} \\
'smooth-debiased'  & -- & -- & -- & -- & 113 & 8.84 \\
'featured-or-disk-debiased' & 44 & 81.2 & 17 & 88.2 & -- & -- \\
'edge-on-debiased'  & -- & -- & 17 & 52.9 & -- & -- \\
\hline
\multicolumn{7}{c}{GZD-5} \\
'smooth-debiased'  & -- & -- & -- & -- & 290 & 38.9 \\
'featured-or-disk-debiased' & 88 & 72.7 & 28 & 78.5 & -- & -- \\
'edge-on-debiased'  & -- & -- & 28 & 53.5 & -- & -- \\
\hline
\end{tabular}
\caption{This table compares the agreement between the redshift debiased vote fractions for morphological types in GZD and the corresponding OZb types. A pair is considered agreeable if it has a likelihood of at least 0.5 (i.e., a majority) for the corresponding morphological type. Here, \textit{N} denotes the total number of overlapping pairs with GZD, while \textit{\%} indicates the fraction of those pairs that are in agreement.}
\label{tab:morphology_comparison_debiased}
\end{table}

\section{Implications of the Expanded Catalogue}
\label{implications}


\textcolor{black}{Expanding and growing the catalogue of occulting pairs is similar in nature to expanding single galaxy catalogues, albeit with one major difference: our targets are on several orders of magnitude rarer than those of single galaxies. Consequently, any growth in this catalogue, however incremental, represents a significant gain in statistical power. Increasing the number of identified pairs directly improves the robustness of our analyses across all pair types, including systems with favourable geometries for dust attenuation and similar studies.}

\textcolor{black}{We summarise the increase in sample size of our catalogue relative to previous occulting-pair catalogues in Table \ref{catalogue_compar}. The growth of the each pair category, while all benefiting from an increase in sample size, also  has distinct implications for different pair types (see Section \ref{paircat} for definitions). For F-type pairs, the expanded catalogue increases the likelihood of identifying pairs with the ideal geometry (e.g., VV191a/b) necessary for a detailed analysis of dust attenuation and related measurements along the outer regions of the spiral disk (see \citealp{Keel13, Keel23, Robertson24, Robertson25}). Some of the pairs identified in the first row of Figure \ref{classification_images} are potentially ideal pairs we could use for such analysis. Similarly, \(\phi\), Q, and X-type pairs also require somewhat ideal geometry for detailed analysis of vertical dust structures in the foreground or background edge-on disk galaxy, and thus benefit from the same statistical gain. S-type pairs are particularly useful for measuring and comparing optical and UV extinction \citep{Keel01b}, and their scientific utility likewise increases with sample size. B-type pairs require considerable care when using these small background sources to map extinction, making larger samples particularly valuable for mitigating individual system uncertainties. Finally, E-type pairs, unlike ideal-geometry pairs, require larger samples to identify low-opacity diffuse dust components. Our catalogue represents a substantial increase over \cite{Keel13} in this front (see Table \ref{catalogue_compar}), so our statistical power is greatly increased for these pairs.}

\textcolor{black}{By increasing our sample size, we are also no longer limited to studies of individual systems, such as what was done with VV191a/b \citep{Keel23, Robertson24, Robertson25}, but can instead conduct large-scale, population-level analyses of these occulting pairs. This enables a systematic investigation of the optical properties (the transmission) of the ISM across diverse galaxy types and environments. This project and its future iterations will serve as the foundation for these analyses. Work toward extending such studies across a wide range of occulting pairs is already underway, following the isophotal and rotational methodologies outlined in \cite{Holwerda09}.}

\begin{table}[ht]
    
    \centering
    \begin{tabular}{ccc}
        \hline
        Pair Type & \multicolumn{1}{c}  {\cite{Keel13}} & {Butrum et al. (2026)} \\
        \hline
        F  & 369 & 42\\
        Q  & 237 & 35 \\
        $\Phi$ & 156 & 48  \\
        X  & 200 & 39  \\
        SE & 102 & 125 \\
        S  & 584 & 110 \\
        B  & 181  & 24  \\
        E  & 59 & 342 \\
        \hline
    \end{tabular}
    \caption{A comparison of pair counts classified in OZb with those in \cite{Keel13}}
    \label{catalogue_compar}
\end{table}


\section{Project Limitations}
\label{limitations}
Some of the discrepancies identified in Section \ref{accuracy} can be attributed to inherent limitations in the project itself. Following the conclusion of the Overlap Zoo beta period, we identified several key issues in our workflow, selection criteria, \textcolor{black}{and image generation process} that likely contributed to these discrepancies. Work is already underway to address these issues and will likely result in significant improvements in upcoming iterations.

\subsection{Workflow Structure}
\label{workflow}

Some aspects of our workflow (Figure \ref{DecisionTree}) are not ideal. For example, responses to Task T00 could better align with the structure used in previous Galaxy Zoo projects.  Our current responses for this task frequently lead to misclassification of edge-on spiral galaxies as ellipticals, which may partly explain discrepancies in classifications. Task T01 may mistakenly label Spiral-Spiral systems as having an elliptical galaxy in the foreground. More importantly, it restricts tasks T06–T09 to only those pairs where users can distinguish between the foreground and background galaxies, which limits our ability to collect annotations for all valid pairs and complicates the querying process. \textcolor{black}{Additionally, Task T01 does not provide information on whether the foreground or background galaxy occupies the central or off-central position. This absence currently makes direct comparison of morphological classifications with GZD-1, -2, and -5 difficult for most pairs, but possible for some (see Section \ref{ozbvsgzd}).} Task T02 also introduces confusion, as users sometimes misidentify Spiral-Elliptical pairs as systems with both galaxies edge-on. Lastly, Task T05 appears too subjective, as the meaning of "significantly smaller" is not clearly defined, warranting caution when using these classifications. This ambiguity is also present in the expert catalogue. We have begun addressing these issues in the next iteration of the project by restructuring the decision tree and removing subjective questions, with the goal of improving consistency and clarity throughout the workflow.

The number of votes per galaxy pair in our workflow is also a significant limitation to consider.  Each occulting pair candidate in our catalogue was retired after receiving just four user classifications (see Section \ref{workflow}). This number is considerably lower than in other citizen science projects such as Galaxy Zoo \citep{Lintott11a, Willett13, Walmsley22}, where each galaxy typically receives upwards of 40 classifications. Due to the relatively low number of votes, we often had to correct for ties in the classifications (see Section \ref{ties}), which may introduce additional uncertainty into the final labels. We anticipate that increasing the number of classifications per image would not only improve the accuracy illustrated in Figure \ref{redshift_accuracy} but also reduce some discrepancies between our work and the expert catalogue. Future public iterations of the project will raise the retirement threshold, allowing more classifications per image and enhancing overall classification reliability.

\subsection{Selection Criteria}
Our image selection criteria could also be improved to better filter out false overlaps (see Section \ref{sampleselect}). For instance, imposing stricter thresholds on the number of votes and vote fractions for galaxies flagged as merging could reduce misclassification of interacting or merging pairs. \textcolor{black}{Unfortunately, we cannot rely on redshift information, as both galaxies in a pair do not always have measured redshifts.} We would also like to impose thresholds to identify false overlaps involving stars; however, this is more challenging. At the current resolution of DESI, stars often appear as tiny, unresolved blobs, making them difficult to differentiate from smooth galaxies, meaning we can't use the 'star or artifact' response in Galaxy Zoo projects to exclude these. Properly separating the two will likely require higher-resolution data from future surveys and visual classifications from additional Galaxy Zoo projects using this data.

\subsection{Image Generation Process}
\label{imagegen}
\textcolor{black}{For the Overlap Zoo Beta project, images were generated as JPEG cutouts from the DESI Legacy Survey. While sufficient for visual inspection, JPEG compression inherently reduces information compared to lossless PNG images. This compression may have introduced minor biases in volunteer classifications, particularly for more distant pairs where subtle morphological features can be blurred due to compression. As a result, the improvements offered by DR10 over earlier releases may have had a limited impact in the current iteration. In the next iteration, we plan to produce images following the methodology of previous Galaxy Zoo projects (e.g., \citealp{Walmsley22}) to avoid this issue entirely.}

\section{Conclusion}
\label{conclusion}

We have presented Overlap Zoo Beta, a beta citizen science project that tests the ability of non-experts to classify occulting galaxy pairs at the imaging depth and quality of \textcolor{black}{DR10 of} the DESI Legacy Survey. \textcolor{black}{The project also presents a catalogue comprising three subset catalogues: (1) an initial 2,380 candidate pairs presented and classified by volunteers, (2) 1,428 total annotations derived from these candidates (including duplicates), and (3) a sample of 765 occulting galaxy pairs with classifications that could be categorised following \cite{Keel13}, as well as associated redshift and magnitude data.} Out of the \textcolor{black}{765 identified} pairs, 
\textcolor{black}{\(\sim17\%\)} have known redshift differences significant enough to eliminate the possibility of interaction \textcolor{black}{(\(\Delta{z} > 0.008\); \citealp{Keel13})}. However, we warrant caution when using these morphological classifications, as each image received only four votes per galaxy pair from non-expert classifiers.

Future work will focus on refining our classification workflow and reclassifying these pairs using a larger number of classifications per galaxy pair, thereby improving classification reliability. We also plan to include the \cite{Keel13} catalogue in the next iteration to take advantage of the improved imaging depth that the DESI DR10 provides over SDSS DR7. Looking further ahead, we aim to apply our classification workflow to data from upcoming large-scale surveys such as Euclid \citep{EuclidQ1}, LSST \citep{Ivezic19}, and the forthcoming Nancy Grace Roman Space Telescope, with support from future Galaxy Zoo projects.

\paragraph{Acknowledgements}
\small{The author(s) would like to thank Mike Walmsley and the Galaxy Zoo team for providing the initial dataset used to create this catalogue. 

The author made use of Grammarly to assist with the drafting of this article to enhance readability and language. After using this tool/service, the authors reviewed and edited the content as needed and take full responsibility for the content of the publication}

\paragraph{Data Availability Statement}
\small{The catalogue generated in this work is available at \textcolor{blue}{\url{https://doi.org/10.5281/zenodo.16884326}}}.

\printendnotes

\bibliography{references, Bibliography}
\appendix

\end{document}